# Optimizing wide-field coded aperture imaging: radial mask holes and scanning

Jonathan E. Grindlay[*a] and JaeSub Hong[a]
[a]Harvard College Observatory, Harvard-Smithsonian CfA, Cambridge, MA 02138

**ABSTRACT**

Imaging at hard X-ray energies (~10-600 keV) over very large fields of view (~60° per telescope) is required to conduct a high sensitivity all-sky and all-time survey for black holes. The proposed Energetic X-ray Imaging Survey Telescope (EXIST) could achieve the high sensitivity required for the mission science objectives by scanning an array of wide-field coded aperture telescopes with aperture mask holes radially aligned to minimize auto-collimation by the thick (~7mm) masks required for high energy imaging. Simulation results from a preliminary design study are reported which quantify the improvement in off-axis imaging sensitivity vs. the conventional case with mask holes all perpendicular to the mask. Such masks can be readily constructed from a stacked laminate of thin (1mm) laser-etched W sheets. An even more dramatic increase in coded aperture imaging sensitivity, and dynamic range, for a realistic telescope and imaging detector with typical systematic errors can be achieved by continuously scanning the field of view of the telescope over the source region to be imaged. Simulation results are reported for detectors with systematic errors 1-10%, randomly distributed but unknown in each detector pixel. For the simplified case of a 1-D coded aperture telescope scanning along its pattern, the systematics are removed identically. Results are also presented for the 2-D case with both 1-D and partial 2-D scanning which demonstrate the feasibility of a coded aperture scanning telescope with systematic errors achieving nearly Poisson-limited sensitivity for signal/background ratios $S/B \sim 10^{-4}$, in constrast to limits typically ~10-100X worse that have been actually achieved by pointed or dithered coded aperture telescopes flown (or proposed) previously.

## 1. INTRODUCTION

Coded aperture imaging, in which a coded mask casts a shadow of a source or collection of sources of arbitrary shape and distribution onto a position-sensitive detector as initially proposed by Dicke[1], allows artifact-free imaging and source timing over very large fields of view providing the mask is occulting (and not scattering). It is thus well-suited to the needs of the proposed all-sky hard X-ray (~10-600 keV) imaging survey for black holes that could be conducted by the Energetic X-ray Imaging Survey Telescope (EXIST), recently described by Grindlay et al[2] and proposed for further study to define the Black Hole Finder Probe in NASA's recently announced Beyond Einstein Program.

The science goals of EXIST are ambitious: to conduct a full-sky survey for black holes (BHs) with $5\sigma$, 1year sensitivity of $S_{min}$(~10-150 keV) ~0.05 mCrab (i.e. $5 \times 10^{-5}$ that of the Crab nebula, the brightest hard X-ray calibration source in the sky). The corresponding energy flux limit $F_{min}$(~10-150 keV) = $5 \times 10^{-13}$ erg cm$^{-2}$ s$^{-1}$ in any band spanning a factor of 2 in energy is comparable to that achieved in the soft X-ray all sky imaging survey conducted with ROSAT[3]. Given the likely Cd-Zn-Te (CZT) imaging detector contemplated[2] for EXIST, the sensitivity in the band ~150-600 keV would be reduced by a factor of ~5-10 due primarily to the detector and mask becoming transparent. In order to achieve such sensitivities as well as relatively large duty cycles (>20%) for exposure on any given source, wide-field (~60° FWHM) imaging and an array of overlapping telescopes are needed. The required coded aperture telescopes then have backgrounds dominated by the cosmic diffuse flux below ~150 keV, with total count rate $B_{min}$(~10-150 keV) ~0.5Crab per unit area-time. The desired EXIST sensitivity limit imposes the stringent requirement that coded aperture imaging be Poisson (not systematics) limited down to signal/background ratios $S/B \sim 10^{-4} - 10^{-3}$, or a factor of ~10-100 below that achieved in previous non-imaging hard X-ray missions such as the narrow-field chopped HEXTE instrument[4] on RXTE or the pointed-dithered coded aperture imagers IBIS[5] now flying on INTEGRAL or the BAT[6] to be launched on Swift.

---

[*] contact author: josh@cfa.harvard.edu; Harvard Observatory, 60 Garden St., Cambridge, MA 02138 (*note: color graphics included*).

In this paper we describe and give preliminary simulation results for two new innovations for coded aperture imaging. The first is to optimize the sensitivity and uniformity of response across the large field of view (FoV) and moderately high energy range and angular resolution desired for EXIST and summarized[2] in the "reference design" by using curved (or faceted) coded aperture masks with mask holes radially aligned across the FoV. The second is to deal with the much more formidable problem of minimizing the effects on imaging sensitivity of systematic errors in the detector - e.g. randomly distributed variations in the gain of each detector pixel that are unmeasurable, or slowly varying, to some fixed level. These will have significant effects on the minimum value of S/B, and thus imaging sensitivity, that can be achieved. The separate problem of systematic effects in the imaging domain (e.g. bright sources on the edge of the FoV and only partially coded; or the effects of coding noise by support structures in the coded aperture mask) are also partly considered. Continual scanning of the coded aperture telescope (array) across the source distribution to be imaged (e.g. the entire sky) is found to have a major positive effect and dramatically reduce the limiting S/B for the detector systematics case; preliminary simulation results for the imaging systematics case shows that scanning is also effective.

## 2. WIDE-FIELD CODED APERTURE IMAGING

Coded aperture imaging has been studied for many applications, and numerous techniques have been developed to optimize the imaging response. Beginning with the introduction of uniformly redundant array (URA) aperture mask patterns with identically flat side-lobe response[7] for which a variety of algorithms exist to construct open vs. closed hole patterns of square, rectangular, or linear format[8], a wide range of construction and analysis techniques have been developed as summarized originally by Caroli et al[9] and recently in a useful website maintained by In't Zand[10]. Nevertheless, the needs for a wide-field (~30-60° FWHM) imager that allows high angular resolution up to the high energy limits imposed by mask occultation (vs. scattering) have not been explicitly considered. Nor have the separate, but complementary and even more crucial requirements imposed by maximizing the dynamic range of S/B, as needed for a wide-field imager to attain maximum sensitivity to very faint sources in the presence of both detector- and image-plane systematics as mentioned above.

The wide-field coded aperture imaging case is shown schematically in Figure 1. The key consideration is the ratio of unit cell diameter d (an individual mask open hole) to its length, given by the mask thickness t, since as shown in Figure 2 this effectively defines a collimator for off-axis imaging. In the low energy limit, where the mask thickness $t \ll d$, this is not important; but for a mask of thickness 7mm and hole size d=2.5mm, as considered for EXIST[2] in order to achieve the required 5arcmin angular resolution with a minimal mask-detector distance of 1.5m, the resulting mask collimation would restrict the imaging FoV to a FWHM auto-collimation angle $\theta_a = atan(d/t) = 20^o$, or significantly less than the ~30° flat-topped and ~60° FWHM value desired. Even if the mask auto-collimation is forced to not exceed the desired fully-coded imaging FoV for a chosen URA, it still imposes an undesirable triangular collimator response across that FoV as shown in Figure 3. Although the clustering of open holes in an actual URA reduces the auto-collimation, as shown in Figure 4, this is not achieved if the coded mask is constructed with a superimposed support grid (which still auto-collimates) as needed to support isolated or `island' closed-hole segments of the mask.

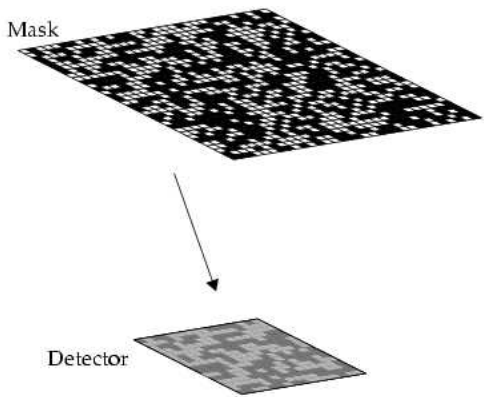

**Figure 1.** Wide-field coded aperture imaging.

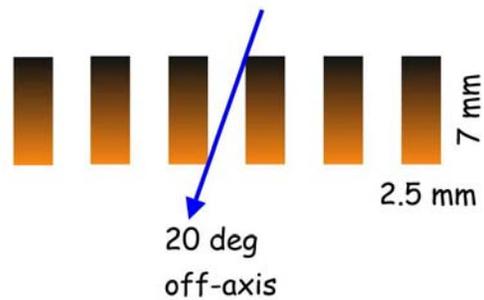

**Figure 2.** Auto-collimation by uniform mask holes of diameter 2.5mm in coded mask of thickness 7mm.

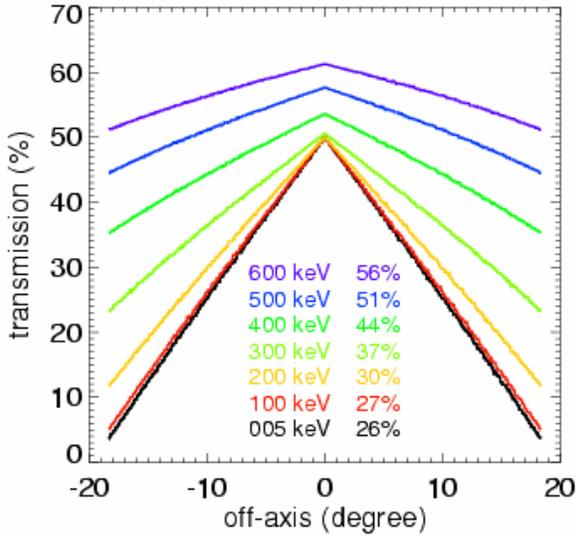 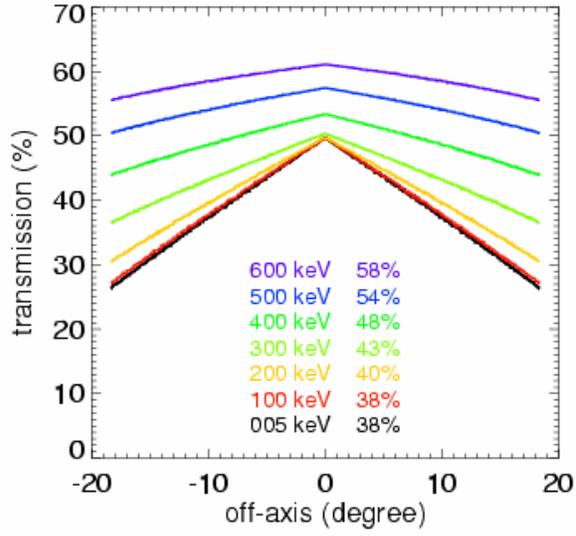

**Figure 3.** Energy-dependent auto-collimation by uniformly spaced mask holes.

**Figure 4.** Auto-collimation imposed by real URA (1-D: 199 element) mask holes.

Auto-collimation could be reduced by increasing the mask hole diameter, d, but with the undesirable consequence that the imaging resolution $\delta\theta = d/f$ for a coded aperture telescope with mask-detector distance f can only be preserved for correspondingly larger f, which increases the overall telescope size and mass which scale roughly as $f^2$. The solution is to impose curvature (or facets) in the coded mask across the FoV and align the holes radially.

## 3. RADIAL HOLE MASKS

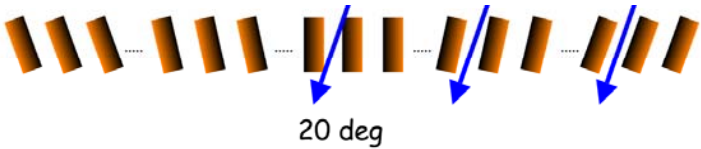

**Figure 5.** Radial mask hole concept (left-most ray is outside FoV).

The radial mask hole concept is shown schematically in Figure 5 for comparison with Figure 2. A possible method for implementation is shown in Figure 6, where the mask is constructed with flat segment facets to follow the approximate curvature of the total FoV and the mask holes are radially aligned smoothly across each facet. The radius of curvature of the mask is chosen to minimize the auto-collimation at the edge of the fully-coded FoV (the left-most ray shown in Figure 5 would be well beyond the fully-coded region). The construction of such a mask could be achieved by laminating the mask from thin laser-etched segments (e.g. for EXIST: 7 segments of W, each 1mm thick) and expanding the scale for each segment as shown in Figure 6. The facets might each be ~50cm x 60cm, to match the size of a ``sub-telescope" detector (see below) for EXIST. They could be supported by a thin frame positioned so as to align underneath the closed row and column segments of the URA mask chosen. Thus, each facet would incorporate an entire URA pattern, and the mask for the fully-coded FoV would then be constructed from 2 x 2 contiguous facets. The non-planar mask segments will introduce small projection effects although these can be minimized by mounting the detector modules on a similarly-faceted array.

An alternative implementation of the radial mask hole concept could employ a smooth quasi-spherical coded mask instead of flat facets. This might be achieved with a (new) variation of hexagonal URA patterns[11], rather than rectangular or square formats. The projection effects may impose more distortions in this case, but have not been studied for their radial-hole implementation.

The auto-collimation properties of the radial-hole URA mask (1-D: 2 cycles of a 199 element URA) are shown in Figure 7 for comparison with the vertical hole case (conventional mask) shown in Figure 4. The ratio of mask throughput between center and edge (18° off axis) is reduced from 47/27=1.63 (Fig. 4) to 45/38=1.18 (Fig. 7) at 100 keV by using radial mask holes, and the overall throughput (percentage numbers included in the figures) is also increased at each energy. The actual imaging performance of the conventional vs. radial hole masks are shown in Figures 8 and 9, which are derived for the same 199 element 1-D URA and show that imaging performance is enhanced over the FoV by radial hole masks. In addition to the future work needed for radial mask design, we are now fabricating a test mask using the flat laminate approach (Fig. 6) for an experimental demonstration (lab and balloon) of the radial hole mask concept.

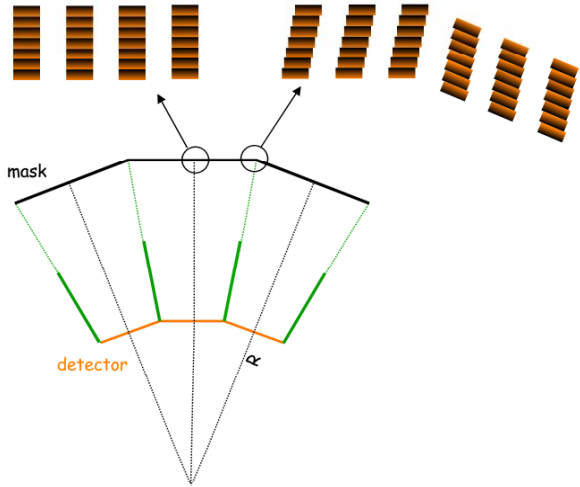
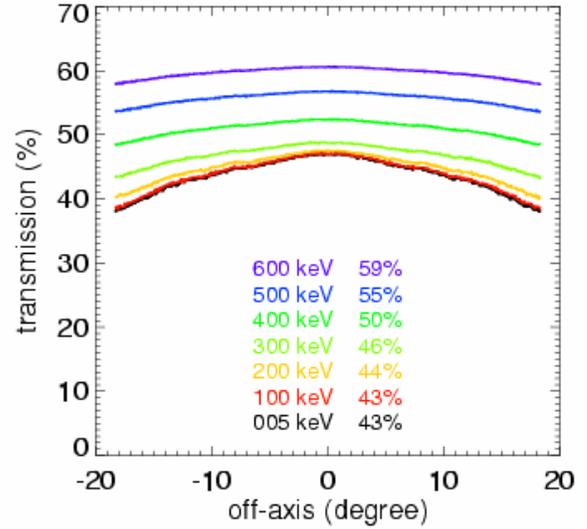

**Figure 6.** Implementation of radial-hole mask with mask facets matched to sub-telescope collimation and laminates.

**Figure 7.** Auto-collimation imposed by radial-hole mask with 290cm radius of curvature.

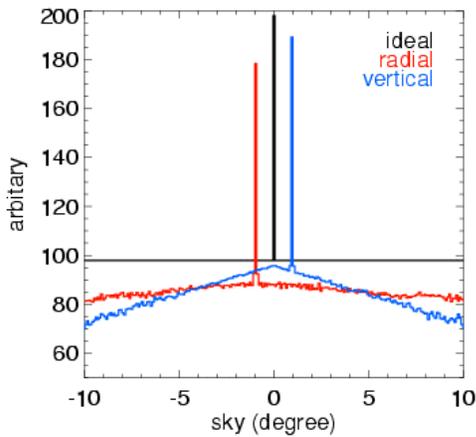
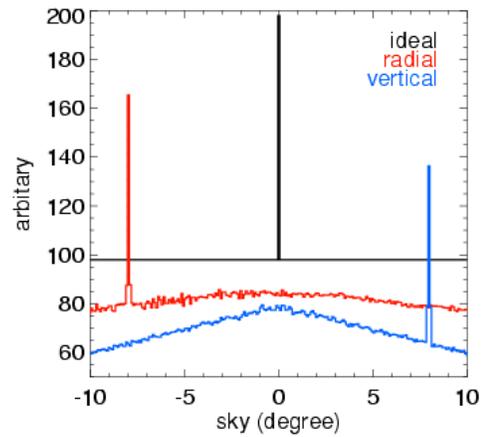

**Figure 8.** 1° off-axis imaging of radial vs. normal masks.

**Figure 9.** 8° off-axis imaging of radial vs. normal masks.

# 4. SCANNING CODED APERTURE IMAGING

Since coded aperture imaging with a URA achieves identically flat side-lobes only if the detector response is also flat and if the detector is uniformly illuminated by the source distribution, departures in either of these produce additional noise in the reconstructed image and reduce the sensitivity achieved. Even if the detector response is well-measured or modeled, it is likely that intrinsic uncertainties in position-sensitive response or gain will always be present, and usually changing (even if only slowly). Although flat-fielding methods[12,13] will partially compensate for background or detector non-uniformities, these are also limited in their usefulness and achievable dynamic range. Although dithering the pointing direction has been used (and is now incorporated on INTEGRAL and will be on Swift), this does not compensate for systematic effects which include multiplicative uncertainties (e.g. gain or response). Similarly, mask vs. anti-mask chopping, as inherently included in rotating hexagonal URAs[11], are primarily effective only for additive uncertainties (e.g. variations in local detector background) and do not correct for unknown multiplicative errors. Thus coded aperture imaging has thus far not demonstrated imaging sensitivities at signal/background levels much below S/B ~ 0.01. As mentioned above, the goals of EXIST require at least an order of magnitude (if not more) improvement. We thus consider the benefits of continually scanning the coded aperture telescope across the FoV to be imaged. The derived benefits of this have motivated the scanning (zenith-pointing, large FoV) of the reference EXIST design.

## 4.1 One-Dimensional imaging and scanning

We begin by considering 1D imaging with a 1D coded aperture constructed with a mask of dimension N = 251 elements and compare the measured signal/noise ratio, SNR, as derived from the image vs. the input value of the SNR expected from Poisson statistics for a source of flux S (cts cm$^{-2}$ sec$^{-1}$) measured in the presence of a background with flux B for an integration time T in a detector with effective area A, which includes geometric and efficiency factors:

$$SNR_{input} = S(A\,T/(S + B))^{0.5}$$

In Figures 10 and 11 we plot the measured SNR, or $SNR_{reconstructed}$ vs. $SNR_{input}$ for imaging with a 1D URA vs. random mask and for the case where the detector elements have randomly distributed systematic errors with mean value 1% (i.e., the counts recorded in any given pixel –whether S or B – have +/-1% variation randomly distributed about their mean value). This simulates a detector with gain values per pixel that can only be measured (or monitored) to an accuracy of ~1% over the time duration of the observation. For each case, we plot (different color points) the cases for a fixed-pointed observation and for two cases of 1D scanning: a partial scan, of 1/4 of the 1D mask across the FoV, and for a full 1D scan. Both cases are for the S/B values appropriate to a ``sub-telescope'' (ST) of the EXIST reference design (9 such STs, each with A = 3000 cm$^2$ and B = 0.2 cts cm$^{-2}$ sec$^{-1}$, constitute *each* of the 3 Telescopes currently considered for EXIST) and a T = 3month integration time, the minimum achieved for any source over a year given the >20-25% duty cycle for exposure. The detector pixel size of 1.25mm is one-half the mask pixel size, so imaging factors (loss of sensitivity due to under-sampling the mask shadow) are not important, and the effects of the systematic errors, $\sigma_{sys}$, can be studied. From Figure 10 it can be seen that a complete scan with a URA removes the systematics completely (and identically, regardless of the value of $\sigma_{sys}$), whereas the 1/4 scan gives SNR reduced by a factor of ~3 and conventional pointing is reduced by a factor of ~30. The effect of $\sigma_{sys}$ = 1% irreducible noise for this pointed URA case (with N=251 elements) is to limit the reconstructed SNR to be only ~3% of its input value, with essentially the same results for the random mask case (Figure 11). The random mask, when fully scanned, only achieves a maximum SNR that is ~20; for larger input SNR, the 1% systematics dominate whereas for smaller input signals, scanning allows the Poisson errors to dominate. The reduction in reconstructed SNR for a partial scan (e.g. the factor of ~3 reduction in reconstructed SNR for the 1/4 scan shown) for either the URA or random mask cases is dependent on 3 variables: N, $\sigma_{sys}$, and the fraction f of the mask scanned. An analytic approximation for this SNR reduction factor, R(N, $\sigma_{sys}$, f), will be given elsewhere.

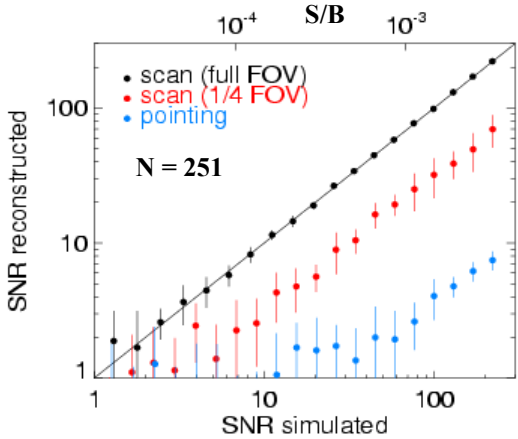
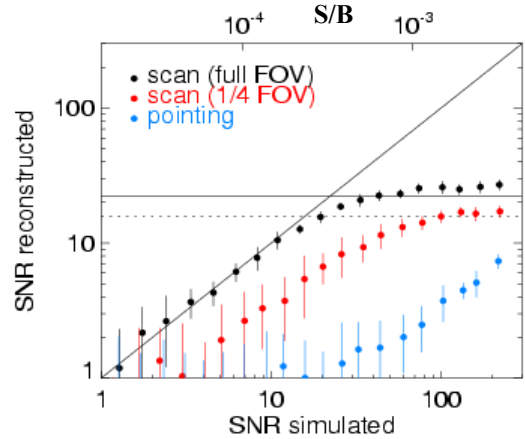

**Figure 10.** Reconstructed vs. input SNR for 1D URA imaging with 1% systematics: pointing vs. scanning.

**Figure 11.** Same as Figure 10 but for random 1D coded mask (again, with N = 251 elements).

### 4.2 Two-Dimensional imaging and 1D vs. 2D scanning

We now consider the 2D imaging case, as needed for EXIST. In Figures 12 and 13 we show the dependence on the mask pixel number, N = M x P, for URA masks of dimension M x P, as well as the effect of full 2D scanning of the mask vs. only a 1D scan of the 2D mask. For each case, the detector parameters (A, B, T) are the same as for Figures 10 and 11, and for each case the mask pattern is extended in a 2 x 2 cycle array so the full field is imaged uniquely. The systematic errors are again randomly distributed in each detector pixel with mean value $\sigma_{sys}$ = 1%. The pointed mask case is also shown for both N values. Once again, for a full 2D scan over the 2D mask, the systematic errors are removed identically (Figure 12). For a 1D scan of the 2D mask, the reduction in reconstructed SNR is by a factor of R ~5 and ~1.1 for the values of N shown in Figures 12 and 13, respectively. The reduction factors for the pointed observation (blue points) are R ~ 20 and 5, respectively.

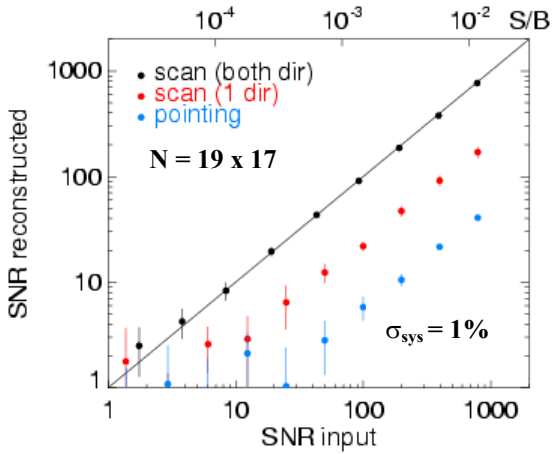 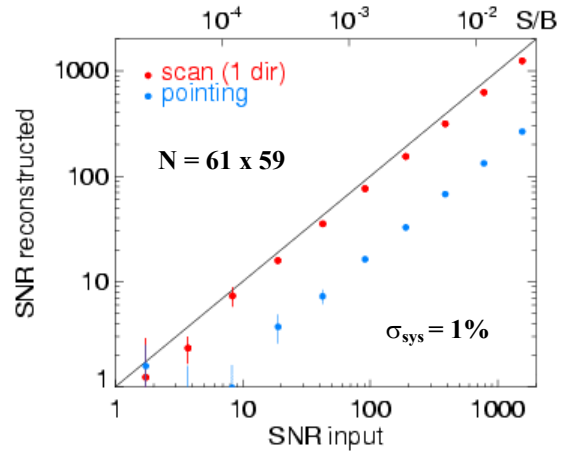

**Figure 12.** 2D URA mask: 2D vs. 1D scanning vs. pointing.  **Figure 13.** 2D URA mask: 1D scanning vs. pointing.

We have simulated the two cases shown in Figures 12 and 13 for an increased value of the systematic noise, with the mean $\sigma_{sys}$ = 10% and again randomly distributed over the detector pixels. The reduction in SNR is then by factors R ~50 and ~10 for the two values of N, respectively, for the 1D scanning case. Thus the reconstructed SNR is reduced approximately in proportion to the value of $\sigma_{sys}$, which increased by a factor of 10. The same result holds for the pointed case. Thus although the reconstructed SNR is improved for larger N, and even 1D scanning of a 2D coded aperture can recover most of the Poisson SNR, the relative increase is limited by the imposed value for the mean systematics error.

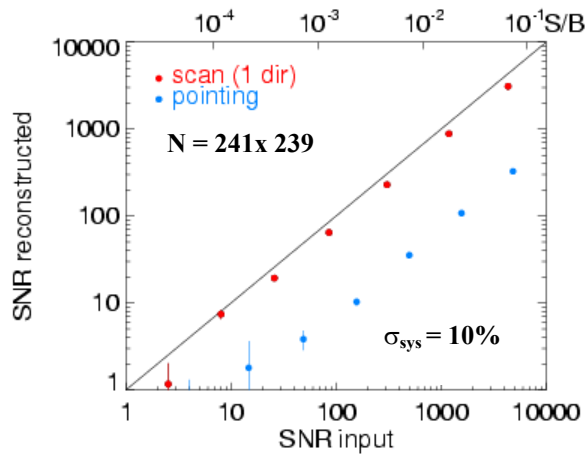

**Figure 14.** 1D scan vs. pointing sensitivity for EXIST.

Finally, we consider a 2D imager scanned in just one direction and with N = 241 x 239 elements in a URA mask as might be appropriate for a single ST on EXIST for which the fully coded FoV of 20° x 25° and 5.7arcmin resolution could be achieved with a 2 x 2 cycle implementation of this URA. Using again the same detector parameters (A,B,T) appropriate for a ST on EXIST, results are shown in Figure 14 for the 1D scanning vs. pointing cases. The reduction factors in SNR from Poisson statistics are R ~ 1.3 and ~15, respectively, for this case where the very conservative case for systematics, with $\sigma_{sys}$ = 10%, is used. This demonstrates that even with a (very) large assumed value for the irreducible systematic errors in the detector, nearly the Poisson SNR can be recovered with just 1D scanning of the 2D URA envisioned for EXIST, and that Poisson-limited imaging sensitivity over the full dynamic range, S/B ~ $10^{-4}$ – 1, is possible. Since each ST for EXIST would complete a full 1D scan across the FoV in only ~5min at the orbital scan rate of ~4° per minute, the requirements for constancy of $\sigma_{sys}$ are very forgiving: systematics need only be constant over a ~5min timescale. The possible sources of such systematics (variations in gain, small changes in detector temperature, etc.) are likely to have much longer time constants that are comparable to (at least) the ~95min orbital period planned for EXIST. We also note that the actual scan track over any source on the sky to achieve the total T = 3month exposure (over ~1y) assumed for the simulation shown in Figure 14 is more nearly 2D since the scan track angle across the detector changes with each satellite orbit. Thus the expected SNR is even closer to the ideal (Poisson) case for the parameters shown.

In Figure 15 we show a simulated high latitude scanning image (T ~3month total exposure per source over ~1y) centered on the bright quasar 3C273 (source A; assumed to have flux 5mCrab) with mean $\sigma_{sys}$ = 10%. The other sources in the field are drawn from the expected logN – logS distribution for the number N (per square degree) of active galactic nuclei (AGN) detected with flux S or greater, $dN/dS \propto S^{-3/2}$, with ~1 AGN deg$^{-2}$ normalization[14] at the ~0.05 mCrab flux threshold for the EXIST survey. Within the 22° x 22° FoV shown around 3C273, some 470 sources are detected above the expected threshold flux of ~0.05mCrab (5σ). The brightest source (5mCrab), 3C273, is detected at ~470σ in the scanning image (left; where in fact 2D scanning is assumed, given the averaging over a wide range of scan angles) but only at ~15σ in the pointed image simulation which is partially shown on the right.

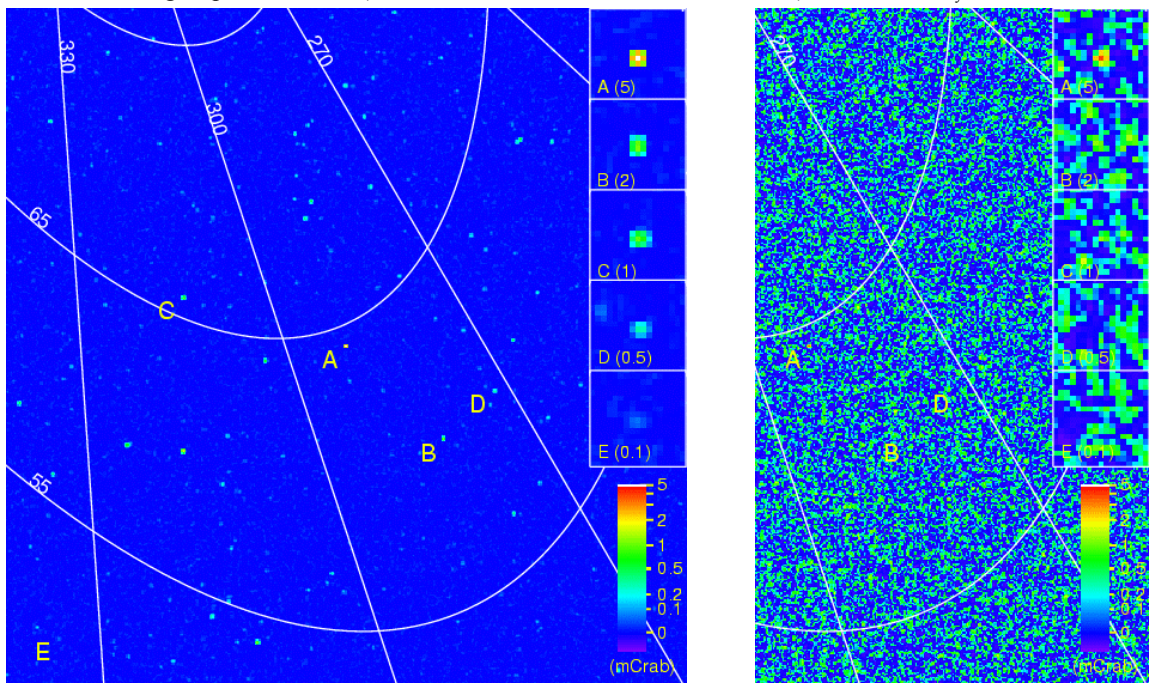

**Figure 15.** Simulated EXIST image around 3C273 (source A) for ~1y scanning (**left**) vs. pointing (**right**; only partial image shown). Fluxes (mCrab) for sources A – E are given in insert image panels along right, with color-bar scale below. Some 470 AGN sources are detected above ~0.05mCrab in this high latitude field (lines of galactic longitude and latitude are shown).

### 4.3 Temporal variability of backgrounds and sources and partial coding efffects

The dramatic improvement to reconstructed SNR enabled by scanning would seem to require that the unknown systematics and also detector backgrounds and source fluxes be constant over the ~5min duration of the scan of any given source across the FoV of a given ST. Whereas this is unlikely to be important for detector systematics ($\sigma_{sys}$), as noted above, particle-induced backgrounds across the detector and certainly bursts and fast transient sources do change on this timescale. In general, particle background induced variations will be smooth across the detector since they are not modulated by the coded mask; and since they are additive (not multiplicative), they are subtracted by the on – off nature of the imaging (sky background pixels on the detector for a given source are increased by the same excess δB background counts as are source pixels). Thus particle-induced background variations over the scan should have a relatively small effect on the reconstructed SNR but more detailed simulations are needed and planned.

Fast timescale variations of cosmic sources, such as a ~10sec gamma-ray burst (GRB), are equivalent to short ``pointings'' and so the scanning cannot be effective. This means the achieved SNR in detecting and localizing the centroid of a GRB image is reduced from a Poisson value by a factor R ~ 3 for a detector with $\sigma_{sys}$ ~ 2%, as might be expected. Again, more detailed simulations are needed and planned. However, this prompt GRB sensitivity (e.g. for real time processing on board) can be partly restored by post-processing where the initial determination of the GRB (or bright source) position allows iterative removal of sources (see Skinner and Connell[15] for a recent implementation) to further enhance the SNR of the source distribution. Similarly, the fast timing properties (e.g. pulsations or bursts) of persistent sources can be fully recovered since back-projection preserves source arrival times.

The longer timescale variability of bright persistent hard X-ray sources (e.g. GRS1915+105 or CygX-1) can also affect the sensitivity for imaging and spectra of much fainter sources within the same FoV since bright source variations are likely to be on timescales (~5min or less) comparable with an individual ST scan. Simulations have been partly carried out to explore this; e.g. source variability with an assumed sinusoidal shape has been introduced, and SNR plots like Figure 14 constructed. Details will be reported elsewhere when a broader range of variable source amplitudes and timescales have been investigated, but preliminary results are encouraging: for a 19 x 17 URA (simple test case) and a purely 1D scan, sinusoidal modulation from (0 – 1) x S over the scan period, and a range of input SNR values from ~3 – 1000, the SNR plot was indistinguishable from the 1D scan (red points) for the constant source case shown in Figure 12.

Finally, as a (moderately) bright and variable source moves through the FoV during the scan, it will become only partly coded when it has moved off the ``flat-top'' (imposed by the diverging collimator; see Figure 6) fully-coded region where a source casts a shadow of the full URA pattern onto the detector and instead only a partial URA shadow is recorded. This will introduce coding noise, since the mask pattern is now effectively (pseudo-) random, and so imaging sensitivity and dynamic range is correspondingly reduced (e.g., see SNR plots for URA vs. random masks as shown in Figures 10 and 11). Thus the partial coding[16] of (moderately) bright sources would degrade the imaging sensitivity for fully-coded faint sources. Fortunately, the EXIST reference design concept allows this to be largely fixed: although a source is indeed only partially coded in a given sub-telescope, the adjacent ST is simultaneously recording the same source at an offset angle (a full Telescope consists of an array of 3 x 3 STs, each with 20° x 25° fully-coded and flat-topped FoV, and mutually offset to give the combined fully-coded and flat-topped FoV of 60° x 75° for the Telescope), so that the combined data from the two STs allows full coding of the source. The price to be paid is that this source is then imaged (for this time period) with two STs and thus the total background B is twice what it would be if fully coded within a single ST. Thus sensitivities are degraded by (only) ~√2. Only the outermost portion of the combined Telescope FoV cannot be corrected in this way, since there is not then an adjacent ST. However even here, for approximately constant sources the imaging can be corrected (fully-coded) by combining successive scans, each of which provides different coverage. Detailed simulations to study all of these effects, as well as the non-uniform detector backgrounds caused by shield leakage and shadowing of telescope support structures, are planned.

## 5. CONCLUSIONS

Wide-field high energy coded aperture imaging, as needed for all-sky survey and monitoring of black holes, can be improved by incorporating coded aperture masks with their aperture holes radially aligned with the wide field of view to be imaged. This flattens the response and improves off-axis imaging sensitivity and resolution. An even more dramatic improvement is gained by having the coded aperture telescope (regardless of its field of view or energy range) execute a continuous scan, as naturally obtained for a zenith-pointing telescope which scans the sky once per orbit. The advantages of scanning are two-fold: it allows systematic errors (e.g. calibration uncertainties in gain or response) to have minimal effect on derived sensitivity and dynamic range; and it allows partial coding effects of off-axis sources to be minimized or largely eliminated, since such sources will be fully-coded for part of the time, by at least part of the telescope array.

This study is based on simulations in progress to optimize the design for the proposed EXIST mission. Further details will be reported, and experimental verification by laboratory and balloon-borne imaging tests are planned using prototype CZT imaging hard X-ray detector arrays with prototype scanning telescopes incorporating radial hole masks.

## ACKNOWLEDGEMENTS

We thank G. Skinner and the EXIST Science Working Group for discussions. This work was supported in part by NASA grant NAG5-5279.